\documentclass{article}
\usepackage{spconf,graphicx,hyperref}
\usepackage{amsmath,epsfig,amssymb,verbatim,amsopn,cite,subfigure,multirow}
\usepackage{balance}
\usepackage{svg}
\usepackage{footnote}
\usepackage{algorithm,algorithmic}
\usepackage{epstopdf}
\usepackage{setspace}
\usepackage{xcolor}
\usepackage{multirow}
\usepackage{url}
\usepackage{array}
\usepackage{pifont}

\usepackage{tikz}

\newcommand{\qe}{{\bf e}}

\newcommand{\qh}{{\bf h}}

\newcommand{\qv}{{\bf v}}
\newcommand{\qw}{{\bf w}}

\newcommand{\qy}{{\bf y}}
\newcommand{\qz}{{\bf z}}

\newcommand{\qH}{{\bf H}}
\newcommand{\qI}{{\bf I}}

\newcommand{\qR}{{\bf R}}

\newcommand{\diag}{\mathrm{diag}}

\newcommand{\Ex}{\mathbb{E}}

\newcommand{\RD}{\mathtt{RD}}
\newcommand{\UB}{\mathtt{UB}}
\newcommand{\UR}{\mathtt{UR}}
\newcommand{\RB}{\mathtt{RB}}

\newcommand{\BD}{\mathtt{BD}}
\newcommand{\BR}{\mathtt{BR}}

\newcommand{\alfaSIs}{\alpha_{\mathtt{SI}}^2}

\newcommand{\Hdl}{\qH^{\dl}}
\newcommand{\bHdl}{\bar{\qH}^{\dl}}
\newcommand{\bHdlT}{\bar{\qH}^{\dl^\dag}}
\newcommand{\Rdl}{{\qR}_{\dl}}

\newcommand{\Hul}{\qH^{\ul}}
\newcommand{\bHul}{\bar{\qH}^{\ul}}
\newcommand{\bHulT}{\bar{\qH}^{\ul^\dag}}
\newcommand{\Rul}{{\qR}_{\ul}}

\newcommand{\C}{\mathbb{C}}

\newcommand{\bAlpha}{\boldsymbol{\alpha}}

\newcommand{\Kd}{K_{\mathtt{dl}}}
\newcommand{\Ku}{K_{\mathtt{ul}}}
\newcommand{\qyul}{\qy^{\mathtt{ul}}}

\DeclareMathOperator{\UEdk}{\mathrm{UE}^{\mathtt{dl}}_{k}}

\DeclareMathOperator{\UEuq}{\mathrm{UE}_{q}^{\mathtt{ul}}}

\DeclareMathOperator{\Rl}{\mathrm{R}_{\ell}}

\newcommand{\qwqdag}{\qw_q^\dag}

\newcommand{\bwq}{\bar\qw_q}
\newcommand{\bvk}{\bar\qv_k}

\newcommand{\CN}{\mathcal{CN}}
\newcommand{\LL}{\mathcal{L}}
\newcommand{\Sn}{\sigma_n^2}

\newcommand{\qHSI}{\qH_{\mathtt{SI}}}
\newcommand{\ul}{\mathtt{ul}}
\newcommand{\dl}{\mathtt{dl}}
\newcommand{\hulq}{\qh_{q}^{\mathtt{ul}}}

\newcommand{\hulqp }{\qh_{q'}^{\mathtt{ul}}}
\newcommand{\hulon}{\qh_{1}^{\mathtt{ul}}}
\newcommand{\hulKu}{\qh_{K_{\ul}}^{\mathtt{ul}}}

\newcommand{\hdlk}{\qh_{k}^{\mathtt{dl}}}
\newcommand{\hdlon}{\qh_{1}^{\mathtt{dl}}}
\newcommand{\hdlKd}{\qh_{K_{\dl}}^{\mathtt{dl}}}
\newcommand{\bhdlon}{\bar{\qh}_{1}^{\mathtt{dl}}}
\newcommand{\bhdlk}{\bar{\qh}_{k}^{\mathtt{dl}}}
\newcommand{\bhdlKd}{\bar{\qh}_{K_{\dl}}^{\mathtt{dl}}}

\newcommand{\bhulon}{\bar{\qh}_{1}^{\mathtt{ul}}}
\newcommand{\bhulq}{\bar{\qh}_{q}^{\mathtt{ul}}}
\newcommand{\bhulKu}{\bar{\qh}_{K_{\ul}}^{\mathtt{ul}}}

\newcommand{\hdlkT}{\qh_{k}^{ {\mathtt{dl}}^\dag}}

\newcommand{\hURBql}{\qh_{q,\ell}^{\mathtt{URB}}}

\newcommand{\D}{\mathtt{D}}

\newcommand{\hUBq}{\qh_{\D,q}^{\ul}}

\newcommand{\hURql}{h_{q,\ell}^{\mathtt{UR}}}
\newcommand{\hRBl}{\qh_{\ell}^{\mathtt{RB}}}

\newcommand{\HRBRl}{\qH_{\ell}^{\mathtt{RBR}}}
\newcommand{\HRBRlp}{\qH_{\ell'}^{\mathtt{RBR}}}
\newcommand{\hBRl}{\qh_{\ell}^{\mathtt{BR}}}
\newcommand{\hBRlT}{\qh_{\ell}^{\mathtt{BR}^T}}
\newcommand{\hRdlk}{h_{\ell k}^{\mathtt{RD}}}

\newcommand{\hBDk}{\qh_{\D,k}^{\dl}}

\newcommand{\hRdlpkc}{h_{\ell' k}^{\mathtt{RD}^*}}
\newcommand{\hURqlpc}{h_{q,\ell'}^{\mathtt{UR}^*}}

\newcommand{\gamakdl}{\gamma_k^{\dl}}
\newcommand{\gamaqul}{\gamma_q^{\ul}}
\newcommand{\cdlk}{c^{\dl}_k(\bAlpha)}
\newcommand{\culq}{c^{\ul}_q(\bAlpha)}

\newcommand{\huRol}{h_{1,\ell}^{\mathtt{UR}}}
\newcommand{\huRkul}{h_{K_u,\ell}^{\mathtt{UR}}}
\newcommand{\roul}{\rho_{\ul}}
\newcommand{\rodl}{\rho_{\dl}}

\newcommand{\alfal}{\alpha_{\ell}}
\newcommand{\alfalsq}{\alpha_{\ell}^2}
\newcommand{\alfalp}{\alpha_{\ell'}}
\newcommand{\alfaln}{\alpha_{\ell}^{(n)}}
\newcommand{\alfalpn}{\alpha_{\ell'}^{(n)}}

\newcommand{\qhURl}{\qh_{\ell}^{\mathtt{UR}}}

\newcommand{\sluq}{s_{q}^{\ul}}
\newcommand{\sluqp}{s_{q'}^{\ul}}

\newcommand{\skdl}{s_{k}^{\dl}}
\newcommand{\skpdl}{s_{k'}^{\dl}}

\newcommand{\Kus}{\mathcal{K}_{\ul}}
\newcommand{\Kds}{\mathcal{K}_{\dl}}

\newcommand{\SINRulq}{\mathrm{SINR}^{\ul}_{q}}
\newcommand{\SINRdlk}{\mathrm{SINR}^{\dl}_{k}}
\newcommand{\SEulq}{\mathrm{SE}^{\ul}_{q}}

\newcommand{\SEdlk}{\mathrm{SE}^{\dl}_{\mathtt{k}}}

\newcommand{\sumL}{\sum\nolimits_{\ell\in\LL}}
\newcommand{\sumLp}{\sum\nolimits_{\ell'\in\LL}}
\newcommand{\sumLpz}{\sum\nolimits_{\ell'\in\{\LL,\mathtt{SI}\}}}
\newcommand{\sumKd}{\sum\nolimits_{k\in\Kds}}

\newcommand{\sumKpdsk}{\sum\nolimits_{k'\in\Kds\setminus k}}
\newcommand{\sumKu}{\sum\nolimits_{q\in\Kus}}
\newcommand{\sumKpu}{\sum\nolimits_{q'\in\Kus}}

\newcommand{\sumQLLp}{ \widetilde{\sum\nolimits}_{q,\ell,\ell'}}

\newcommand{\SEulQ}{\mathcal{S}_{\ul}}
\newcommand{\SEdlQ}{\mathcal{S}_{\dl}}
\newcommand{\TTdl}{T_{\dl}}
\newcommand{\tdl}{t_{\dl}}
\newcommand{\TTdln}{T_{\dl}^{(n)}}

\newcommand{\TTul}{T_{\ul}}
\newcommand{\TTuln}{T_{\ul}^{(n)}}
\newcommand{\tul}{t_{\ul}}

\graphicspath{{Figures/}}
\title{Repeater-Assisted Massive MIMO Full-Duplex Communications}
\name{
\fontsize{0.34cm}{1cm}\selectfont Mohammadali Mohammadi$^\dagger$, Dhanushka Kudathanthirige$^\star$, Himal A. Suraweera$^\ddagger$, Hien Quoc Ngo$^\dagger$, and Michail Matthaiou$^\dagger$\thanks{The work of H. Q. Ngo was supported by the U.K. Research and Innovation Future Leaders Fellowships under Grant MR/X010635/1 and by a research grant from the Department for the Economy Northern Ireland under the US-Ireland R\&D Partnership Programme. The work of M. Matthaiou was supported by the European Research Council (ERC) under the European Unions Horizon 2020 research and innovation programme (grant agreement No. 101001331).}}
\address{\fontsize{0.37cm}{1cm}\selectfont
$^\dagger$Centre for Wireless Innovation (CWI), Queen's University Belfast, U.K.\\
\fontsize{0.37cm}{1cm}\selectfont
$^\star$School of Engineering, Macquarie University, Australia\\
\fontsize{0.37cm}{1cm}\selectfont
$^\ddagger$Department of Electrical and Electronic Engineering, University of Peradeniya, Sri Lanka\\
}
\begin{document}
%
\maketitle
\begin{abstract}
We consider a wireless network comprising multiple single-antenna repeaters that amplify and instantaneously re-transmit received signals in a full-duplex (FD) communication setting. Specifically, we study a massive multiple-input multiple output base station that simultaneously serves multiple uplink (UL) and downlink (DL) user equipment (UE) over the same frequency band. The focus is on the  problem of repeater weight optimization at each active repeater to maximize the sum of the weighted minimum spectral efficiencies (SEs) for both UL and DL UEs. The resulting non-convex optimization problem is tackled using a successive convex approximation technique. To demonstrate the effectiveness of the proposed approach, we evaluate its performance against benchmark systems with and without repeater assistance. The optimized FD design achieves SE improvements of up to $4$‑fold and $2.5$‑fold compared to its half‑duplex counterpart.
\end{abstract}
\vspace{-0.2em}
%
\vspace{-1em}
\section{Introduction}~\label{sec:intro}
Massive multiple-input multiple-output (mMIMO) is central to 5G and beyond, yet its practical performance and coverage are often limited by signal attenuation, line-of-sight (LoS) propagation, and shadowing~\cite{Hien:TCOM:2013}. These limitations have prompted research on distributed MIMO networks, including network MIMO, coordinated multipoint, and cell-free massive MIMO~\cite{Mohammadi:JSAC:2023}. However, such systems require tight synchronization and heavy fronthaul infrastructure, which inevitably amplify their implementation complexity. Alternatively, network-controlled repeaters, acting as rich scatterers with amplification, can be deployed to unlock performance gains and enhance the coverage~\cite{Carvalho2025}.

Few papers have studied repeater-assisted (RA) MIMO systems in terms of fundamental limits, design, and optimization. In \cite{Tsai:TWC:2010}, the capacity scaling and coverage with fixed-location repeater deployment under LoS conditions were studied. In \cite{Willhammar:MCM:2025}, RA MIMO was evaluated against two other MIMO architectures to highlight its signal-to-interference-plus-noise ratio (SINR) enhancement. In \cite{Iimori:GC:2023} and \cite{topal2025fair}, a repeater amplification strategy and multiple activation control algorithms were proposed, to improve system performance and enable energy-efficient operation. The maximum amplification gain that repeaters can use without causing system instability due to positive feedback from inter-repeater interference was investigated in \cite{Larsson:SPAWC:2024}. The findings in~\cite{bai2025repeater} demonstrated that repeaters can considerably enhance system performance in both sub-6 GHz and millimeter-wave bands.  

Despite recent progress, gaps remain in the literature. Existing works assume HD operation, whereas FD terminals can support simultaneous UL and DL transmissions with high SE~\cite{Mohammadi:TCOM:2015,Mohammadi:TWC:2018}, which makes it an attractive solution for beyond 5G systems~\cite{smida_jsac_2023}.  In this paper, we develop a framework to quantify the performance gains enabled by repeater swarms in multiuser FD mMIMO systems. We analyze both DL and UL performance in terms of SE, considering zero-forcing (ZF) precoding and combining at the FD base station (BS). To mitigate the additional interference introduced by repeaters, such as noise injection and signal retransmission, we propose an efficient iterative algorithm for optimizing repeater gains with the objective of maximizing the weighted sum of the minimum UL and DL SE. Numerical results show notable performance gains over benchmark schemes, reflecting the combined impact of repeaters, FD operation, and optimized amplification gains. In terms of fairness, the proposed design achieves more than a $10$‑fold SE improvement compared to conventional non‑repeater‑assisted mMIMO systems.

\textit{Notation:} We use bold lower (capital) case letters to denote vectors (matrices). The superscript $(\cdot)^\dag$ stands for the Hermitian operation. A circular, zero-mean, symmetric complex Gaussian distribution having variance $\sigma^2$ is denoted by $\mathcal{CN}(0,\sigma^2)$. Finally, $\mathbb{E}\{\cdot\}$ denotes the statistical expectation.  
\vspace{-1.5em}
\section{System Model}~\label{sec:systemmodel}
Consider a RA mMIMO FD system  that simultaneously serves $\Kd$ DL and $\Ku$ UL single-antenna UEs, aided by $L$ surrounding single-antenna repeaters. 
It is assumed that  the mMIMO BS is equipped with $M$ antennas, where $M_t$ antennas are allocated for DL transmission and $M_r$ antennas are dedicated to UL reception. For simplicity of notation, we define the sets $\Kus\triangleq\{1,\ldots,\Ku\}$, $\Kds\triangleq\{1,\ldots,\Kds\}$, and $\LL\triangleq\{1,\ldots,L\}$ to represent the sets of UL UEs, DL UEs, and repeaters, respectively. Moreover, $\Rl$, $\UEdk$, and $\UEuq$ denote the $\ell$-th repeater, the $k$-th DL UE, and the $q$-th UL UE, respectively.

Following the setup in~\cite{Iimori:GC:2023}, the system is assumed to be synchronized, with the repeaters centrally controlled by the network. These repeaters forward the DL (UL) signals to (from) the DL (UL) UEs from (to) the mMIMO BS without causing inter-symbol interference—i.e., the forwarded signals fit entirely within the guard interval. Moreover, the repeaters are assumed to be spatially distributed across the coverage area of the BS, with their positions fixed (e.g., mounted on lampposts)~\cite{Willhammar:MCM:2025}. 

\vspace{-0.5em}
\subsection{Channel Model}
The compound DL channel between the BS and $\UEdk$ is expressed as $\hdlk=\hBDk + \sumL\alfal \hRdlk  \hBRl$, where $\hBDk\in\C^{M_t\times1}$ denotes the direct channel between the BS and $\UEdk$, with entries distributed as $\CN(0, \beta_{k}^{\BD})$. The term $\sum_{\ell\in\LL}\alfal \hRdlk \hBRl$ represents the aggregate indirect channel through the repeaters, where $\alfal$ is the weight applied at the $\Rl$; $\hRdlk\sim\CN(0, \beta_{\ell k}^{\RD})$ denotes the channel between $\Rl$ and $\UEdk$; and $\hBRl\in\C^{M_t\times 1}$ denotes the channel vector between the BS and $\Rl$, with entries distributed as $\CN(0, \beta_{\ell}^{\BR})$.

The compound UL channel between the $\UEuq$ and the BS is denoted by $\hulq\in\C^{M_r\times1}$, which can be expressed as $\hulq =  \hUBq+\sumL\alfal\hURql \hRBl$, where $\hUBq\in\C^{M_r\times 1}$ is the direct channel from the $\UEuq$ to the BS with $\CN(0, \beta_{q}^{\UB} )$ distributed entries; $\hURBql\triangleq\alfal\hURql \hRBl\in\C^{M_r\times 1}$ denotes the channel from the $\UEuq$ through the $\Rl$ to the BS, where $\hURql\sim\CN(0, \beta_{q\ell}^{\UR})$ is the channel coefficient between the $\UEuq$ and $\Rl$, while $\hRBl\in\C^{M_r\times 1}$  denotes the channel vector between the $\Rl$ and the BS, whose entries are distributed as $\CN(0, \beta_{\ell}^{\RB})$.

Moreover, $\qHSI\in\C^{M_r\times M_t}$ represents the self-interference (SI) channel matrix at the BS, i.e., the channel between the BS’s transmit and receive arrays.\footnote{We model the SI channel via the Rayleigh fading distribution, under the assumptions that any LoS component is efficiently reduced by antenna isolation such that the major effect comes from scattering~\cite{Mohammadi:TWC:2018}.} Finally, $h_{qk}\sim\CN(0,\beta_{qk})$ represents the channel between $\UEuq$ and $\UEdk$.

\vspace{-0.5em}
\subsection{DL Signaling} 
Let $\skdl$, with $\Ex\{|\skdl|^2\}=1$, denote the symbol transmitted by the BS towards $\UEdk$, and let $\rodl$ denote the transmit power allocated by the BS. Moreover, we assume that the BS employs a precoding vector $\qv_k \in \C^{M_t\times 1}$ for $\UEdk$, normalized such that $\Ex\{\Vert\qv_k\Vert^2\}=1$. As in~\cite{topal2025fair}, for simplicity, we will disregard the interactions between the repeaters. Then, the received signal at $\UEdk$ can be expressed as
\vspace{-0.5em}
\begin{align}~\label{eq:dlsignal:UEk}
    y_k^{\dl} & = 
     \sum\nolimits_{k'\in\Kds}\sqrt{\eta_{k'} \rho_{\dl}}\hdlkT\qv_{k'}\skpdl
     \nonumber\\
     & + \!\sqrt{\roul} \sumKu
    \Big(
    h_{q k}
    \!+\!
    \sumL    
    \alfal \hRdlk \hURql  
    \Big)\sluq 
    \nonumber\\
    & +
    \sumL
    \alfal \hRdlk n_{\mathtt{R},\ell} + n_k,
\end{align}
where $0\!\leq\!\eta_k\!\leq\! 1$ is the power allocation coefficient for $\UEdk$, $\rho_{\dl}$ is the transmit power at the BS, $\sluq$, with $\Ex\{\vert \sluq\vert\}=1$, denotes the symbol transmitted by $\UEuq$ towards the BS and $\roul$ represents the transmit power by the $\UEuq$, $n_{\mathtt{R},\ell}\sim\CN(0,\Sn)$ and $n_k\sim\CN(0,\Sn)$ denote the additive white Gaussian noise (AWGN) at $\Rl$ and $\UEdk$, respectively.

Therefore, the SE of the $\UEdk$ is given by  $\SEdlk = \log_2(1+\SINRdlk)$, where $\SINRdlk$ denotes the received SINR at $\UEdk$, as defined in~\eqref{eq:SINR:dlk}, at the top of the next page.
\begin{figure*}[t]
\normalsize
\vspace{-1.5em}
\begin{align}\label{eq:SINR:dlk}
    \SINRdlk 
    \!=\!
    \frac
    {{\eta_k \rho_{\dl}}\big\vert
    \hdlkT\qv_k
    \big\vert^2}
    {\rho_{\dl}   
    \sumKpdsk
     \eta_{k'} 
     \big\vert \hdlkT
    \qv_{k'}\big\vert^2
    \!+\!
    \roul
    \sumKu
    \!\!
    \Big(
    \vert h_{q k}
    \!+\!
    \sumL
    \alfal \hRdlk \hURql\vert^2  
    \Big)
    \!+\!\Sn
    \Big( 
    \sumL
    \vert\alfal \hRdlk \vert^2\!+\!1\Big)}
\end{align}
\vspace{-1.5em}
\end{figure*}


\vspace{-0.5em}
\subsection{UL Signaling}
The estimate of the data symbol $\sluq$ at the BS is given by $y_{q}^{\ul}=\qwqdag\qyul$, where $\qw_q\in\C^{M_r\times 1}$ with $\Ex\{\Vert \qw_q\Vert^2\}=1$ denotes the received combining vector used to decode the data from $\UEuq$, while
\vspace{-1em}
\begin{align}~\label{eq:yul}
   \qyul &= \sqrt{\roul}\sumKpu\hulqp \sluqp
   \nonumber\\
    &+ {\alpha_{\mathtt{SI}}}\qHSI\sumKd\sqrt{\eta_k \rho_{\dl}} \qv_k \skdl
    \nonumber\\
    &+\sumL\!
    \sumKd\!
    \sqrt{\eta_k \rho_{\dl}}\alfal \HRBRl \qv_k\skdl\nonumber\\
    &
    +
    \sumL\!\alfal \hRBl n_{\mathtt{R},\ell},+\qz,
\end{align}
where the second term represents the SI at the BS, the third term corresponds to the received copy of the DL signal relayed through the repeaters, and the fourth term denotes the amplified copy of the noise received at the repeaters. In~\eqref{eq:yul},  $\HRBRl=\hRBl\hBRlT\in\C^{M_r\times M_t}$ is the loopback channel through the $\Rl$, where $\hRBl\in\C^{M_r\times 1}$  denotes the channel vector between the $\Rl$ and BS, whose entries are distributed as $\CN(0, \beta_{\ell}^{\BR})$, while $\qz\sim\CN(\boldsymbol{0},\Sn \qI_{M_r})$ is the AWGN vector at the BS with variance $\Sn$ per each element. We assume that the SI can be mitigated to some extent using a combination of analog- and digital-domain cancellation techniques at the BS, as described in~\cite{smida_jsac_2023}. The residual SI is characterized by the parameter $0 \leq \alpha_{\mathtt{SI}} \leq 1$, which quantifies the effectiveness of SI cancellation at the BS.

Therefore, the SE of the $\UEuq$ is given by  $\SEulq= \log_2(1+ \SINRulq)$, where $ \SINRulq$ denotes the received SINR at $\UEuq$, as defined in~\eqref{eq:SINR:ulk}, at the top of the next page.

\begin{figure*}[t]
\normalsize
\vspace{-1.5em}
 \begin{align}\label{eq:SINR:ulk}
    \SINRulq 
    &\!=\!\frac
    {\roul\big\vert\qwqdag  \hulq\big\vert^2}
    {  
    {\roul}
    \sum\nolimits_{q'\in\Kus\setminus q}
    \big
    \vert\qwqdag \hulqp\big\vert^2 
    \!+\! 
    \sumKd
    \sumLpz    
    \!{\eta_k \rho_{\dl}}\big \vert\alpha_{\ell'}\qwqdag\HRBRlp\qv_k \big\vert^2 \!+\!\Sn
    \big(
    \sumL
    \!\big\vert \alfal \qwqdag\hRBl\big\vert^2 \!+\!\big\Vert\qwqdag\big\Vert^2\big)}
\end{align}  
\vspace{-1.2em}
 \hrulefill
\end{figure*}

\vspace{-0.5em}
\subsection{ZF Precoding and Combining}
Let $\Hdl \!=\! [\hdlon\!,\ldots,\hdlKd]$ and $\Hul\!=\![\hulon\!,\ldots,\hulKu]$. Then, given that $M_t$ and $M_r$ are large, we can express these channel matrices as $\Hdl = \bHdl\Rdl^{1/2}$ and $\Hul = \bHul\Rul^{1/2}$, where $\bHdl = \![\bhdlon,\ldots,\bhdlKd]$ and $\bHul = [\bhulon,\ldots,\bhulKu]\!\in \!\C^{M_r\times K_{\ul}}$ with $\bhulq\in\C^{M_r\times 1}$, $\bhdlk\in\C^{M_t\times 1}$ and $\bhulq\in\C^{M_r\times 1}$  containing zero-mean and unit variance elements. Moreover,  $\Rdl = \diag\{\gamma_1^{\dl},\ldots, \gamma_{K_{\dl}}^{\dl}\}\in\mathbb{R}^{K_{\dl}\times K_{\dl}}$ and and $\Rul = \diag\{\gamma_1^{\ul},\ldots, \gamma_{K_{\ul}}^{\ul}\}\in\mathbb{R}^{K_{\ul}\times K_{\ul}}$, where  $\gamakdl = \beta_k^{\mathtt{BD}} + \sumL \alfalsq \beta_{\ell k}^{\mathtt{UR}} \beta_{\ell}^{\mathtt{BR}}$ and $\gamma_q^{\ul} = \beta_q^{\mathtt{UB}} + \sumL \alfalsq \beta_{q\ell }^{\RD} \beta_{\ell}^{\mathtt{BR}}$. Then, we apply the ZF principle to design $\qv_{k}$ and $ \qw_{q}$ as
\vspace{-0.8em}
\begin{align}
\qv_{k} = 
\frac{\cdlk}{\sqrt{\gamakdl}} 
\bvk
\quad \text{and} \quad
\qw_{q} =
\frac{\culq}{\sqrt{\gamaqul}} 
\bwq,
\label{eq:vkwq}
\end{align}
respectively, where $\bvk\triangleq\bHdl \big (\bHdlT \bHdl\big) ^{-1} \qe_{v_k}$, $\bwq =
\bHul \big (\bHulT \bHul\big) ^{-1} 
\qe_{w_q}$, while  $\qe_{v_k}$ ( $\qe_{w_q}$) is the $k$-th ($q$-th) column of  $\qI_{K_{\dl}}$ ($\qI_{K_{\ul}}$); $\cdlk $ and $\culq $ are the normalization constant, chosen to satisfy $\Vert \qv_k\Vert^2=\Vert \qw_q\Vert^2=1$. That is, $\cdlk = \sqrt{\gamakdl}/\Vert \bvk \Vert$ and $\culq  = \sqrt{\gamaqul}/\Vert \bwq\Vert$.

\vspace{-0.6em}
\section{System Design Optimization}
From~\eqref{eq:dlsignal:UEk} and~\eqref{eq:yul}, we observe that repeaters inject additional noise and interference in both UL and DL transmissions. This raises the question: to what extent can performance still improve despite these impairments? To address this, we formulate an optimization problem that tunes the repeater weights to maximize the weighted sum of the minimum SE across all DL and UL UEs, while ensuring that the quality-of-service (QoS) requirements of all UEs are met. Mathematically, 
\vspace{-0.7em}
\begin{subequations}\label{P1:opt:maxmin}
\begin{alignat}{2}
&(\mathcal{P}1):~\!\max_{\bAlpha}      
&\hspace{1em}&\omega_{\dl}\min_{k\in\Kds}\SEdlk+\omega_{\ul}\min_{q\in\Kus}\SEulq\label{eq:P1:opt:maxmin:obj}\\
&\hspace{3.5em}\text{s.t.} 
&         & \SEdlk \geq \SEdlQ, \forall k\in\Kds,\label{eq:P1:opt:maxmin:ct2}\\
&         &      &\SEulq \geq \SEulQ, \forall q\in\Kus,\label{eq:P1:opt:maxmin:ct1}\\
&         &      &0\!\leq\!\alfal\!\leq\! \min\Big(\alpha^{\max}\!\!, \sqrt{\frac{P_{\max}}{\Psi_{\ell}}} \Big),
\forall \ell\in\LL,\label{eq:P1:opt:maxmin:ct4}
\end{alignat}
\end{subequations}
where $\omega_{\ul}$ and $\omega_{\dl}$ are the positive weighting factors of the UL and DL parts of the objective function; $\bAlpha\!=\![\alpha_1,\ldots,\alpha_L]^T$;   $\Psi_{\ell} \triangleq \roul\Vert \qhURl \Vert^2 +\rho_{\dl}\sum\nolimits_{k\in\Kds}\eta_k\vert \hBRlT\qv_k \vert^2+\Sn$, while $\qhURl=[\huRol,\ldots,\huRkul]$. Constraints~\eqref{eq:P1:opt:maxmin:ct2} and ~\eqref{eq:P1:opt:maxmin:ct1} denote the QoS requirements, where $\SEdlQ$ and $\SEulQ$ are the minimum SE requirements for DL and UL UEs, respectively. Constraint~\eqref{eq:P1:opt:maxmin:ct4} imposes an upper limit on each repeater’s output power, denoted by $P_{\max}$, while accounting for the maximum amplification factor, $\alpha^{\max}$, which depends on hardware limitations and inter-repeater distance to ensure stability~\cite{topal2025fair}.

By introducing two auxiliary variables $\tul$ and $\tdl$, we recast ($\mathcal{P}1$) as 
\vspace{-1.2em}
\begin{subequations}\label{P2:opt:maxmin}
\begin{alignat}{2}
(\mathcal{P}2):     
&\max_{\tdl,\tul,\bAlpha}      
&\hspace{0.5em}&\omega_{\dl}\tdl+\omega_{\ul}\tul\label{eq:P1:opt:maxmin:obj}\\
&\hspace{1em}\text{s.t.}
&      & \SINRdlk \geq \TTdl, \forall k\in\Kds,\label{eq:P2:opt:maxmin:ct2}\\
&          &         & \SINRulq \geq \TTul, \forall q\in\Kus,\label{eq:P2:opt:maxmin:ct1}\\
&          &         & \TTdl \geq 2^{\tdl}-1, \quad \TTul \geq 2^{\tul}-1,\label{eq:P2:opt:maxmin:ct4}\\
&         &          & \tdl \geq \SEdlQ, \quad \tul \geq \SEulQ,\label{eq:P2:opt:maxmin:ct5}\\
&         &          &\eqref{eq:P1:opt:maxmin:ct4}.\label{eq:P2:opt:maxmin:ct6}
\end{alignat}
\end{subequations}

The optimization problem~ is non-convex, due to the non-convexity of ~\eqref{eq:P2:opt:maxmin:ct2} and~\eqref{eq:P2:opt:maxmin:ct1}. To address this challenge, we propose an amplification control algorithm based on the convex–concave procedure (CCP). However, for given values of $\SEdlQ$ and $\SEulQ$, constraint~\eqref{eq:P2:opt:maxmin:ct5} may lead to infeasibility when the CCP algorithm is initialized with a randomly chosen starting point. To overcome this, we adopt a feasible-point pursuit approach within the CCP framework by introducing nonnegative feasibility parameters, $\varphi_k^{\dl}$ and $\varphi_k^{\ul}$, for each constraint in~\eqref{eq:P2:opt:maxmin:ct2} and~\eqref{eq:P2:opt:maxmin:ct1}, respectively.  Relative regularization terms are then added to the objective function, which iteratively drives these parameters to zero~\cite{Mehanna:SPL:2015}. This guarantees a feasible initialization of the algorithm, while infeasibility of the final solution can be identified through the convergence behavior of the feasibility parameters. For the simplicity of the notation, we define $\sumQLLp\triangleq\sumKu\sumL\sumLp$. Accordingly, we rewrite~\eqref{eq:P2:opt:maxmin:ct2} as
\vspace{-0.4em}
\begin{align}~\label{eq:dl:CCP}
    & \sumL \!
    \frac{\alfalsq}{\TTdl} 
    \xi_1
    \!+\!\sumQLLp\!
   (\alfal\!-\!\alpha_{\ell'})^2 \xi_2
     \!+\!\varphi_k^{\dl} \!\geq\! 
    \sumL
     \alfal^2 \xi_3  
     \nonumber\\
     &\hspace{0em}       
       +\sumKu\!
       \sumL\!\!
    \alfal
    \xi_4  
    \!+\!
    \sumQLLp\!
   (\alfal\!+\!\alpha_{\ell'})^2 \xi_2
    \Big) \!+\! \xi_5,  
\end{align}
\begin{figure*}[t]
\vspace{-0.5cm}
    \centering
    \begin{minipage}[t]{0.32\textwidth}
        \centering
         \includegraphics[trim=0 0cm 0cm 0cm,clip,width=1.1\textwidth]{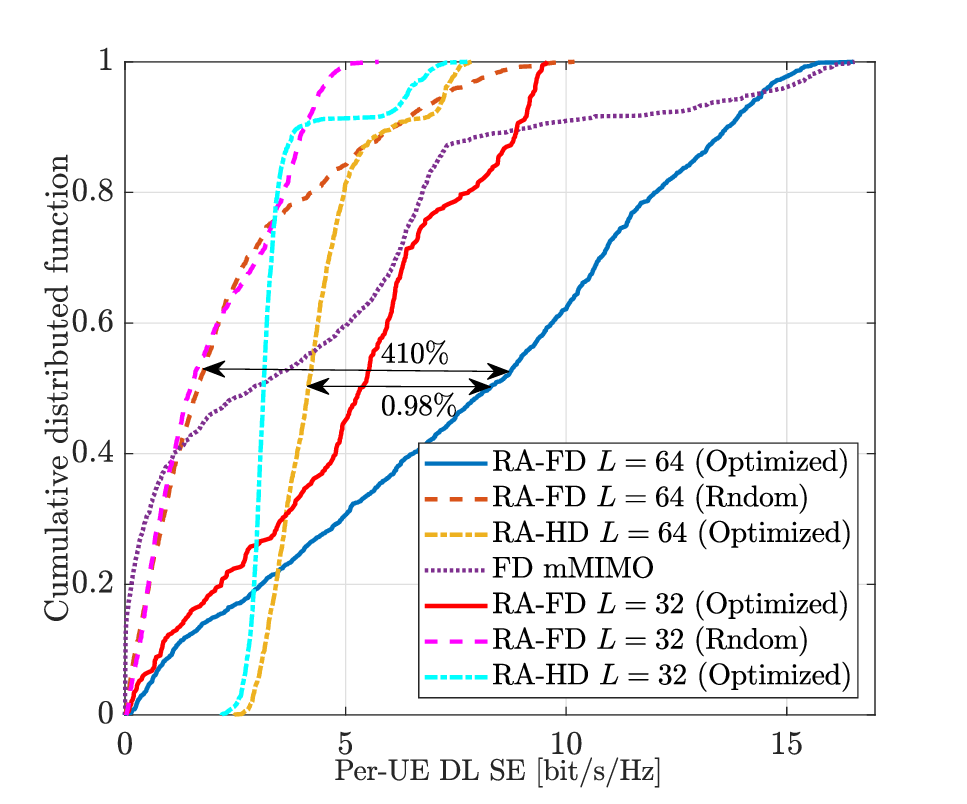}
         \vspace{-1.7em}
        \caption{\small The CDF of the DL per-UE SE for different system designs.}
        \label{fig:DL}
    \end{minipage}
    \hfill
    \begin{minipage}[t]{0.32\textwidth}
        \centering
          \includegraphics[trim=0 0cm 0cm 0cm,clip,width=1.1\textwidth]{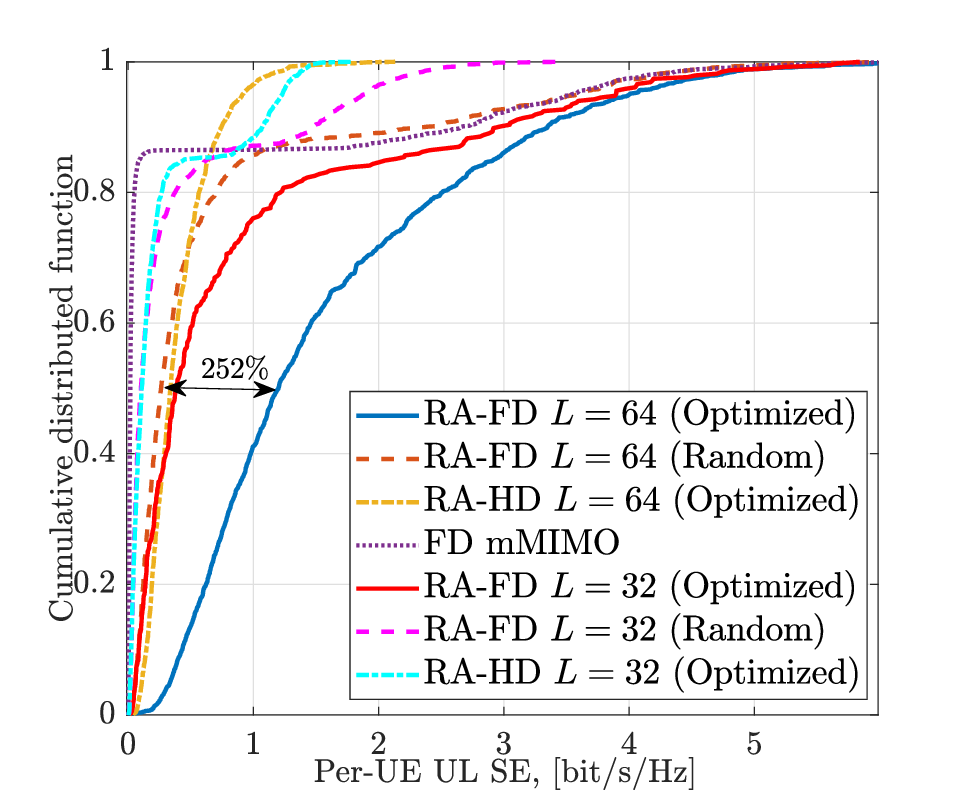}
          \vspace{-1.7em}
        \caption{\small The CDF of the UL per-UE SE for different system designs.}
        \label{fig:UL}
    \end{minipage}
    \hfill
    \begin{minipage}[t]{0.32\textwidth}
        \centering
         \includegraphics[trim=0 0cm 0cm 0cm,clip,width=1.1\textwidth]{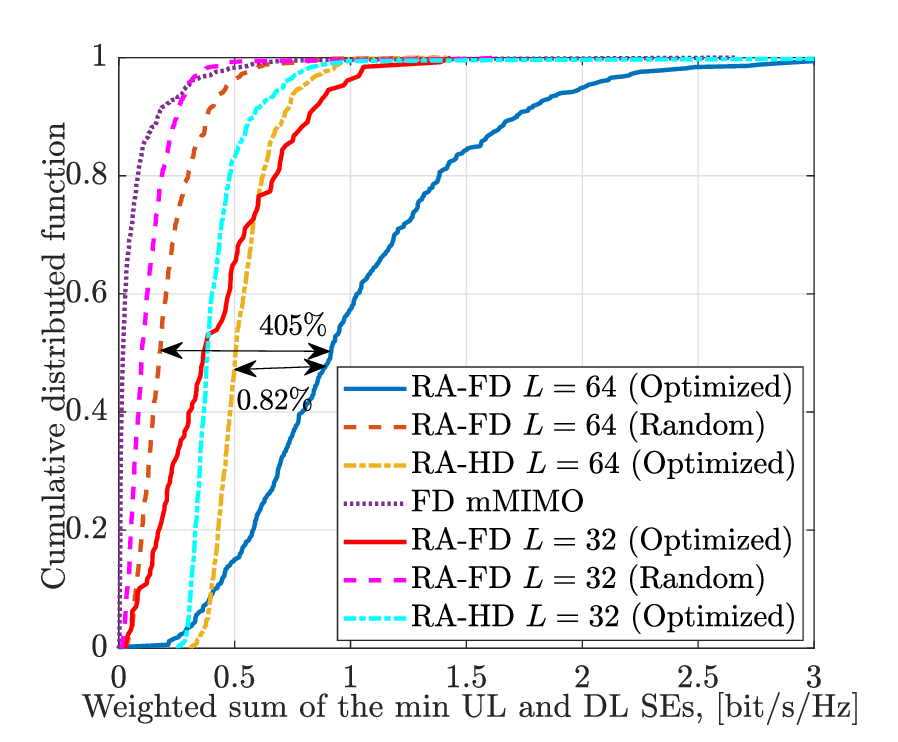}
         \vspace{-1.7em}
        \caption{\small The CDF of the objective function in ($\mathcal{P}1$) for different system designs.}
        \label{fig:sum}
    \end{minipage}
\vspace{-1.1em}
\end{figure*}
where $\xi_1\triangleq\frac{\eta_k \rho_{\dl}}{\Vert \bvk \Vert^2}\beta_{\ell k}^{\UR} \beta_{\ell}^{\BR}$; $\xi_2\triangleq \frac{\roul}{4} \hRdlk \hRdlpkc\hURql\hURqlpc$; $\xi_3\triangleq\Sn\vert\hRdlk \vert^2$; $\xi_4\triangleq 2\roul\Re\{ h_{q k}^*\hRdlk \hURql\}$; $\xi_5\triangleq \Sn -\frac{\eta_k \rho_{\dl}}{\Vert \bvk \Vert^2} \beta_k^{\BD}+ \rho_{\ul}\sumKu\vert h_{q k}\vert^2 $. We notice that~\eqref{eq:dl:CCP} is non-convex due to the presence of a convex term $\frac{\alfalsq}{\TTdl}$ with $\TTdl>0$  and quadratic form at the left-hand side of the inequality. To facilitate the description, we use a superscript ($n$) to denote the value of the involving variable produced after ($n-1$) iterations ($n \geq 0$). In light of successive convex approximation (SCA), constraint~\eqref{eq:dl:CCP} can be approximated by the following convex one
\vspace{-1.1em}
\begin{align}~\label{eq:SEQoSdl:SCA}
    & \sumL\! 
    \frac{\alfaln}{\TTdln}
    \Big(2\alfal\!-\!\frac{\alfaln}{\TTdln}\TTdl\Big)
    \xi_1
    \!+\!\varphi_k^{\dl}
    \!\!-\!\!\sumL\!\!
     \alfal^2 \xi_3  
    \nonumber\\
    &
    +\sumQLLp
   \big(\alfaln\!-\!\alfalpn\big)\Big( 2\big(\alfal\!-\!\alfalp\big) - \alfaln\!+\!\alfalpn\Big) \xi_2
     \geq  
     \nonumber\\
     &\hspace{1em}       
     \sumKu\!\!
     \sumL
    \alfal
    \xi_4  
    \!+\!\sumQLLp
   (\alfal\!+\!\alfalp)^2 \xi_2
    \!+\!
     \xi_5.  
\end{align}

Now, in order to deal with the non-convex constraint~\eqref{eq:P2:opt:maxmin:ct1}, we first define  $\mu_1\triangleq \frac{\Vert \bwq\Vert^2}{\roul}\big( \rodl\sumKd\frac{ \vert \bwq^\dag  \HRBRl \bvk \vert^2}{\Vert \bwq\Vert^2\Vert \bvk\Vert^2}+ \frac{\Sn\vert \bwq^\dag \hRBl\vert^2}{\Vert \bwq\Vert^2}\big)$, $\mu_2 =  \frac{\Vert \bwq\Vert^2}{\roul} \big( \rodl \alfaSIs\sumKd\frac{ \vert \bwq^\dag  \qHSI \bvk \vert^2}{\Vert \bwq\Vert^2\Vert \bvk\Vert^2} + \Sn \big) $. Then, by applying SCA, we approximate ~\eqref{eq:P2:opt:maxmin:ct1} with the following convex constraint
\vspace{-0.9em}
\begin{align}~\label{eq:SEQoSul:SCA}
    &\sumL  \frac{\alfaln}{\TTuln}
    \Big(2\alfal\!-\!\frac{\alfaln}{\TTuln}\TTul\Big) \beta_{k\ell }^{\RD} \beta_{\ell}^{\BR} + \frac{\beta_q^{\UB}}{\TTul}+\varphi_k^{\ul}
  \nonumber\\
  & \geq  
    \sumL
     \alpha_{\ell}^2
     \mu_1 
   +\mu_2. 
\end{align}

To this end, the fairness problem  ($\mathcal{P}1$) can be stated as follows 
\vspace{-0.7em}
\begin{subequations}\label{P3:opt:maxmin:Final}
\begin{alignat}{2}
(\mathcal{P}3):     
&\max_{\tdl,\tul, \varphi_k^{\dl}, \varphi_q^{\ul}, \bAlpha}      
&\hspace{0.5em}&\omega_{\dl}\tdl+\omega_{\ul}\tul 
+ \lambda_{\dl}\sumKd \varphi_k^{\dl}
\nonumber\\
&&&\hspace{1em} + \lambda_{\ul}\sumKu \varphi_q^{\ul}\label{eq:P3:opt:maxmin:obj}\\
&\hspace{2.5em}\text{s.t.}
&      &~\eqref{eq:SEQoSdl:SCA},~\eqref{eq:SEQoSul:SCA},~\eqref{eq:P2:opt:maxmin:ct4},\eqref{eq:P2:opt:maxmin:ct6}\label{eq:P3:opt:maxmin:ct2}
\end{alignat}
\end{subequations}
where $\lambda_{\dl}$ and $\lambda_{\ul}$  are the regularization coefficient. The problem ($\mathcal{P}3$) is a convex optimization problem, which can be efficiently solved in an iterative manner via CVX~\cite{cvx}. 

The problem \eqref{P3:opt:maxmin:Final} involves $A_v\triangleq (L+K_{\dl}+K_{\ul}+2)$ real-valued scalar variables, $A_l\triangleq L$ linear constraints, $A_q\triangleq K_{\dl}+K_{\ul} $ quadratic constraints. Therefore, the algorithm for solving problem \eqref{P3:opt:maxmin:Final} requires a complexity of $\mathcal{O}(\sqrt{A_l+A_q}(A_v+A_l+A_q)A_v^2)$ in each iteration~\cite{tam16TWC}.

\vspace{-0.3em}
\section{Numerical Results and Discussion}~\label{sec:num}
We consider a system with $L=32$, and $64$ repeaters and $\Kd=\Ku = 5$ users, randomly deployed within a $400$\,m$^2$ square area. The BS, located at the center, is equipped with $M_t = M_r = 32$ antennas. Transmit powers and path loss models follow \cite[Table 1]{Iimori:GC:2023}, and for FD operation, the SI attenuation is set to $\alpha_{\mathtt{SI}} = -60$\,dB. 

For performance comparison, we consider the following key MIMO architectures: i) \textbf{\textit{RA-FD (Random)}}: The same system setup as the proposed design but applies randomly generated weights at distributed repeaters. ii) \textbf{\textit{RA-HD (Optimized)}}: The HD counterpart of the proposed system, under the ``antenna preserved" condition~\cite{Mohammadi:TCOM:2015}, is considered, where repeater weights are randomly assigned. In this case, all BS antennas ($M_t + M_r$) are used for transmission and reception by equally sharing the time between the two operations. iii) \textbf{\textit{FD-mMIMO}}: A fully centralized FD system equipped with mMIMO arrays at the BS, without any aids from  repeaters.




We evaluate the performance of the proposed RA mMIMO FD system, where the $\alpha_l$s are optimized according to ($\mathcal{P}3$). We compare RA-FD (optimized) against other MIMO baselines in Fig.~\ref{fig:DL} and~\ref{fig:UL} in terms of the cumulative distribution function (CDF) of the per-UE DL and UL SE. As shown in these two figures, the proposed RA-FD system consistently outperforms the centralized FD mMIMO baseline in both UL and DL scenarios, highlighting the benefit of incorporating distributed repeaters. Furthermore, FD operation yields substantial performance gains in both DL and UL transmissions, with respective improvements of 
$98\%$ and  $252\%$ in the average per-UE SE. Finally, we observe that optimizing the repeater weights is critical, as it enables effective interference management and leads to performance improvements of up to $4$-fold and $2.5$-fold in the RA-FD scenario.

Figure~\ref{fig:sum} shows the CDF of the weighted sum of the minimum UL and DL SEs for different network designs. The results highlight the remarkable gains achieved by incorporating repeaters in the considered network scenario. However, we also observe that if RA-FD is not carefully designed, its performance may fall below that of RA-HD due to the considerable interference and amplified injected noise.

\vspace{-0.1em}
\section{Conclusion}~\label{sec:conc}
We investigated the role of swarm repeaters in RA mMIMO FD networks, demonstrating promising performance gains. By optimizing repeater weights through an iterative algorithm while guaranteeing QoS levels for both DL and UL UEs, we showed that the detrimental effects of injected noise and interference from repeater retransmissions can be effectively mitigated. Consequently, in RA mMIMO systems, FD operation at the BS yields up to an $82\%$ improvement in the system’s minimum weighted sum SE compared to its HD counterpart.

\vfill\pagebreak

\bibliographystyle{IEEEtran}
\bibliography{strings,references}

\end{document}